\documentclass{aa}

\usepackage{amsmath}
\usepackage{graphicx}
\usepackage{psfig}
\usepackage{txfonts}
\usepackage{textcomp}

\def\arcsec{\hbox{$^{\prime\prime}$}}
\def\degr{\hbox{$^{\circ}$}}

\newcommand{\etal}{et~al. }

\newcommand{\caii}{Ca\,{\sc ii}}

\newcommand{\SP}{{Solar Phys.}}

\newcommand{\Ha}{${\rm H\alpha}$ }
\newcommand{\ha}{${\rm H\alpha}$ }

\begin{document}

\title{Transmission and conversion of magnetoacoustic waves on the magnetic canopy in a quiet Sun region}

       \author{I.Kontogiannis\inst{}
        \and G.\,Tsiropoula\inst{}
        \and K.\,Tziotziou\inst{}}

\institute{Institute for Astronomy, Astrophysics, Space Applications and Remote Sensing,
National Observatory of Athens,
GR-15236 Penteli, Greece\\
        \email{[jkonto; georgia; kostas]@noa.gr}}

\offprints{G.\,Tsiropoula\\ \email{georgia@noa.gr}}

\date{Received <date> / Accepted <accepted>}

\abstract{We present evidence for the conversion and transmission of wave modes on the magnetic flux tubes that constitute mottles and form the magnetic canopy in a quiet Sun region.}
{Our aim is to highlight the details and the key parameters of the mechanism that produces power halos and magnetic shadows around the magnetic network observed in \ha.}
{We use our previous calculations of the magnetic field vector and the height of the magnetic canopy, and based on simple assumptions, we determine the turning height, i.e., the height at which the fast magneto-acoustic waves reflect at the chromosphere. We compare the variation of 3, 5, and 7\,min power in the magnetic shadow and the power halo with the results of a two-dimensional model on mode conversion and transmission. The key parameter of the model is the attack angle, which is related to the inclination of the magnetic field vector at the canopy height. Our analysis takes also into account that  1) there are projection effects on the propagation of waves, 2) the magnetic canopy and the turning height are curved layers, 3) waves with periods longer than 3\,min only reach the chromosphere in the presence of inclined magnetic fields (ramp effect), 4) mottles in \ha are canopy structures, and 5) the wings of \ha contain mixed signal from low- and high-$\beta$ plasma.}
 {The dependence of the measured power on the attack angle follows the anticipated by the two-dimensional model very well. Long-period slow waves are channeled to the upper chromospheric layers following the magnetic field lines of mottles, while short-period fast waves penetrate the magnetic canopy and are reflected back higher, at the turning height.} 
{Although both magnetoacoustic modes contribute to velocity signals, making the interpretation of observations a challenging task, we conclude that conversion and transmission of the acoustic waves into fast and slow magnetoacoustic waves are responsible for forming power halos and magnetic shadows in the quiet Sun region.}


\titlerunning{transmission and conversion of waves in a quiet Sun region}
\authorrunning{Kontogiannis \etal}
\maketitle

\section{Introduction}
The chromosphere is probably the most elusive layer of the solar atmosphere. In quiet regions, its complexity is mainly associated with the small scale photospheric magnetic flux concentrations that outline the boundaries of supergranules. These concentrations constitute the magnetic network, a web--like pattern of bright points that, in turn, represent the cross-sections of roughly vertical flux tubes. These magnetic flux tubes expand with height (because the ambient gas pressure drops) and form the so--called magnetic canopy (Gabriel \cite{gab76}). This critical layer marks the transition between the gas--pressure-dominated (below the canopy) and the magnetically dominated atmosphere. The plasma-$\beta$ parameter (with $\beta$ defined as the ratio of the gas pressure to the magnetic pressure) is used to distinguish between these two regimes, and the condition $\beta = 1$ defines the location of the magnetic canopy. In and above the magnetic canopy layer, flux tubes become highly inclined, and the chromosphere in \ha appears as a mesh pattern of elongated dark structures, most of them related to mottles (see Tsiropoula \etal \cite{tsirop12} for a comprehensive review on their properties).

Since the discovery of the 5\,min oscillations by Leighton \etal (\cite{leighton}) and the observation of the p-mode spectrum through ``$k-\omega$'' diagrams by Deubner (\cite{deub75}) and Deubner \etal (\cite{deub79}), several studies have focused on wave propagation in the solar atmosphere (Mein \& Mein \cite{mein76}, Lites \& Chipman \cite{lites79}, Lites \etal \cite{lites82}, Kneer \& von Uexk\"{u}ll \cite{kneer85}, Deubner \& Fleck \cite{deub89}, Fleck \& Deubner \cite{fleck89}, Deubner \& Fleck \cite{deub_fleck90}, Deubner \etal \cite{deub90}, Lites \etal \cite{lites93}). It is widely accepted that wave propagation is allowed for waves with periods shorter than 3\,min (frequencies larger than 5.2\,mHz), known as the acoustic cut-off period, and this means that waves with longer periods do not propagate at greater heights in the chromosphere and are, instead, evanescent. This conclusion is supported by the small phase differences measured between long-period oscillations at different heights. 

Although this general picture would be completely compatible with a homogeneous environment, the existence of strong magnetic concentrations results in substantial deviations. Thus, in the presence of inclined and relatively strong magnetic fields, longer period waves may leak to the chromoshere (Michalitsanos \cite{mich73}, Bel \& Leroy \cite{bel}, Suematsu \cite{suematsu}). This is due to the lowering of the acoustic cut-off frequency, leading to the existence of magneto-acoustic portals (Jefferies \etal \cite{jefferies06}) that channel the long-period waves into the upper atmosphere (De Pontieu \etal \cite{depontieu04}). As a result, the oscillatory properties of the bright points and the chromospheric mottles differ, in general, from those of the internetwork (which is dominated by 3--min oscillations), with periods around 5-7\,min or more being a common finding (Tsiropoula \etal \cite{tsirop09}). 

The intense study of oscillations over extended magnetized areas on the Sun in the past two decades has led to the discovery of power halos and magnetic shadows. Initially detected in the photosphere around active regions (Braun \etal \cite{braun92}, Brown \etal \cite{brown92}, Hindman \& Brown \cite{hindman98}, Braun \& Lindsey \cite{braun99}, Thomas \& Stanchfield \cite{thomas00}) and then also found around the magnetic network in the quiet Sun (Krijger \etal \cite{krijger}, Kontogiannis \etal \cite{kont10a}, Paper I hereafter), power halos are areas where the acoustic power is enhanced. Magnetic shadows, on the other hand, are areas of reduced acoustic power, observed at the chromosphere, over and around the magnetic network (Judge \etal \cite{judge}, McIntosh \& Judge \cite{mcintosh01}, McIntosh \etal \cite{mcintosh03}, Vecchio \etal \cite{vecchio}). In our previous studies (Paper I), we showed that magnetic shadows in the \ha chromosphere around the magnetic network are replaced by power halos in the photosphere. Furthermore, we showed (Kontogiannis \etal \cite{kont10b}, Paper II hereafter) that both phenomena depend on the location of the magnetic canopy and concluded that they are closely related to the interaction of the acoustic waves with the magnetic field that consitutes the mottles and the magnetic canopy.

The interaction of the wave modes with the magnetic field has been addressed by sophisticated numerical simulation studies of increasingly realistic cases, which try to describe the complicated environment of the solar atmosphere. Simulations have shown that at the internetwork, away from strong magnetic fields, upwardly propagating high--frequency acoustic waves shock at the chromosphere and produce the internetwork bright grains in \caii\ H and K (Carlsson \& Stein, \cite{cs97}). The interaction of acoustic waves with the magnetic canopy, which leads to wave mixing and interference, is examined by Rosenthal \etal (\cite{ros02}) and Bogdan \etal (\cite{bog03}) in simple two-dimensional cases. These simulations have demonstrated that the attack angle, defined by the inclination of the magnetic field vector and the direction of the wavevector, is a critical parameter for the interaction between the acoustic waves and the magnetic canopy (Carlsson \& Bogdan \cite{carlsson06}).

Wave propagation, wave mode transformation, the role of the magnetic canopy, and the magnetic field inclination in solar magnetic structures have been examined in several studies (Khomenko \& Collados \cite{khom06}, Khomenko \etal \cite{khom08}, Khomenko \& Collados \cite{khom09}). Schunker \& Cally (\cite{schunker06}) and Cally (\cite{cally07}) used ray theory to describe wave propagation and transformation in two dimensions by combining analytical solutions of the MHD equations with principles of geometrical optics. They distinguished between two processes: transmission and conversion. A wave that undergoes transmission preserves its nature (acoustic or magnetic), but changes character from fast to slow and vice versa. These authors give an analytical expression for the transmission coefficient $T$ that characterizes the efficiency of the process, which is favored by low frequency and small attack angles. On the other hand, in conversion, the fast (or slow) wave changes nature (e.g., from acoustic to magnetic) but remains a fast (or slow) wave. The corresponding coefficient, $C$ is larger when the frequency of the incident waves is higher and the attack angle is larger. We note that $T\,+\,C\,=\,1$, so that energy conservation is satisfied.

Using MHD simulations, Nutto \etal (\cite{nutto10}) studied wave propagation and reported good agreement with the results of ray theory, as described above. In a series of papers, the same authors (Nutto \etal \cite{nutto12a}, \cite{nutto12b}) described the interaction in detail between magnetoacoustic waves and the magnetic field of the network in highly realistic configurations and explained the formation of the magnetic shadows. On the other hand, Stangalini \etal (\cite{stang11}) used the results of ray theory to explain the power distribution of 3 and 5\,min oscillations above an active region, introducing a modified transmission coefficient to take projection effects and p-mode leakage into account. They found very good agreement between the observed and the theoretically predicted power distributions.

The present study is an effort to link these analytical and numerical results to measurable quantities from observations. To this end, we extend our previous work on oscillations in the quiet Sun to seek evidence of conversion and transmission of wave modes on the magnetic flux tubes that constitute mottles and form the magnetic canopy.

\section{Observations}

\begin{figure}[htp]
\centering
 \includegraphics[width=9cm]{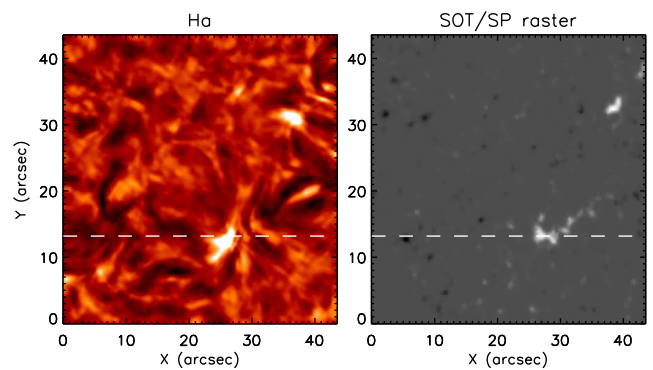}
  \caption[]
    {Image of the FOV in the \ha line center (left) and the corresponding part of the SOT/SP magnetogram (right). The dashed horizontal line marks the position of the vertical slice shown in the right panel of Fig.~\ref{fig:3dplot}.}
    {\label{fig:data}}
\end{figure}

We use observations obtained on October 15, 2007 as part of an observational campaign that included several ground--based and space--born instruments. In this work we utilize data from the Dutch Open Telescope (DOT, Rutten \etal \cite{rutten2004}) and the Spectropolarimeter of the Solar Optical Telescope (SOT/SP, Tsuneta
\etal \cite{sot}) onboard \textit{Hinode}. DOT provided time series of high resolution speckle reconstructed images in five wavelengths along the \ha profile, i.e. at line center, at $\pm$0.35\,{\AA} and  $\pm$0.75\,{\AA} from line center. The cadence of the observations is 30\,s, the duration is 30\,min, and the spatial size is 0.109$\arcsec$. Out of the 84$\arcsec\,\times\,$87$\arcsec$ field--of--view (FOV) of the \ha observations, an area of 44$\arcsec\,\times\,$44$\arcsec$ was selected (Fig.~\ref{fig:data}, left panel). We use the wing intensities in \ha to calculate the Doppler signal (DS hereafter, see Tsiropoula 2000) at $\pm$0.35\,{\AA} and $\pm$0.70\,{\AA} from line center. 

The SOT/SP performed two raster scans of the same area in the 6301.5 and 6302.5\,{\AA} Fe\,I lines at 9:05 and 9:15 UT. The FOV of the raster scan was 50\arcsec$\,\times\,$164$\arcsec$ and the spatial scale of the scans 0.32$\arcsec$ (Fig.~\ref{fig:data}, right panel). 
The HAO/CSAC team (http://www.csac.hao.ucar.edu/csac/dataHostSearch.jsp) produced the corresponding magnetogram by performing the inversion of Stokes spectra via the MERLIN code.
Further details on the observations, the speckle reconstruction procedure of the \ha observations, preliminary reduction steps, and the inversion of the Stokes spectra can be found in Papers I and II and in Kontogiannis \etal (\cite{kontogiannis11}).

\begin{figure*}[htp]
\centering
  \includegraphics[width=6cm]{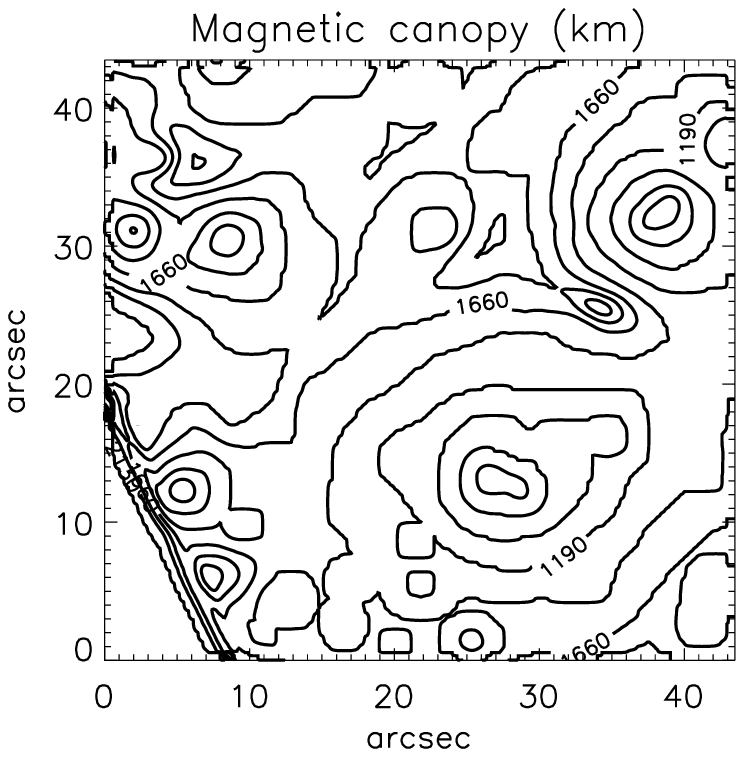}   
  \includegraphics[width=6cm]{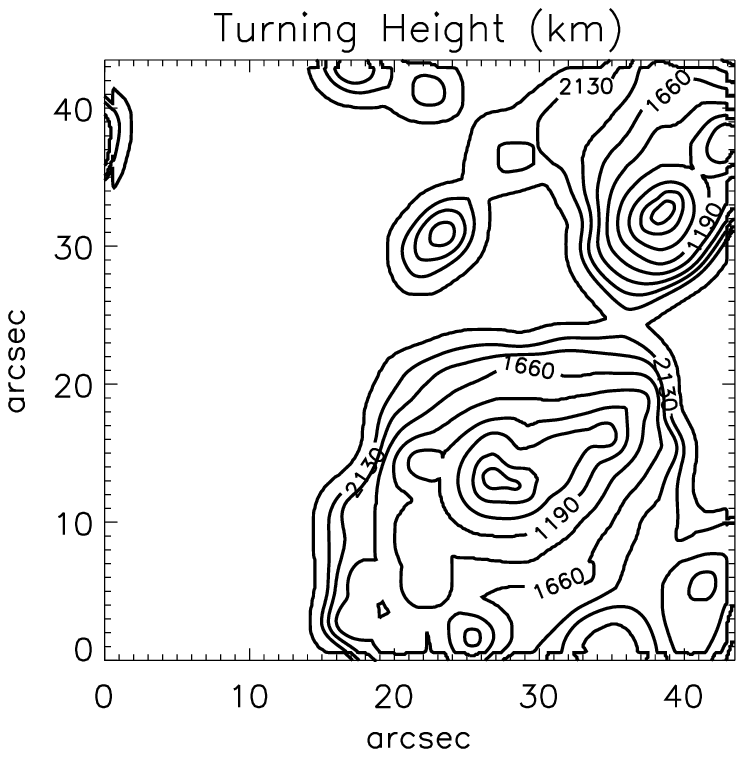}
  \includegraphics[width=5.9cm]{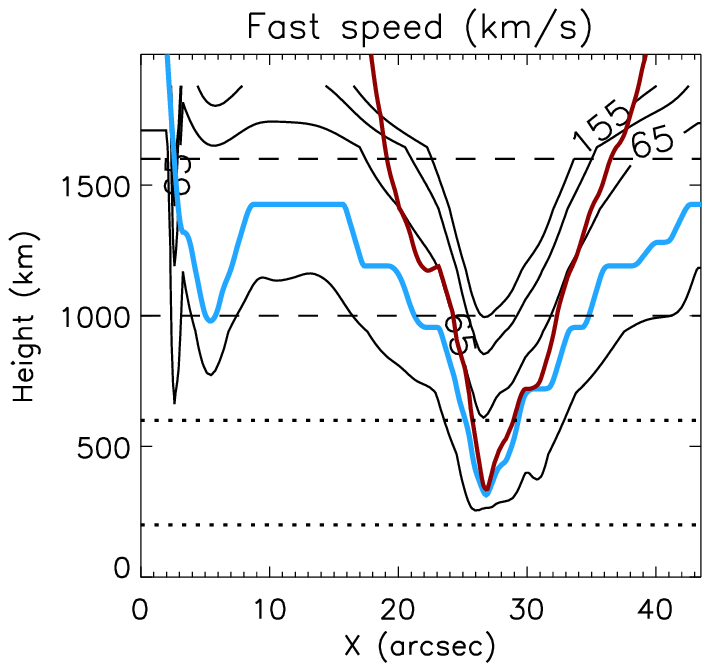}
  \caption[]
    {Contours of the magnetic canopy height and the turning height over the entire FOV shown in Fig.~\ref{fig:data} (left and middle panels, respectively). In the right panel the canopy and turning heights along the dashed line of Fig.~\ref{fig:data} (blue and red lines, respectively) are plotted. These surfaces were smoothed over 1000\,km for better visualization. Overplotted are contours of the fast speed up to a height of 2000\,km. For reference, the estimated HOF (based on Leenaarts \etal \cite{leenaarts}) of $\pm0.70\,\AA$ and $\pm0.35\,\AA$ \ha wings are also plotted (horizontal dotted and dashed lines, respectively).}
    {\label{fig:3dplot}}
\end{figure*}

\section{Analysis}

A study of oscillations in the observed region was carried out using wavelet analysis (Torrence \& Compo, \cite{torr}). This powerful tool gives a two-dimensional spectrum that also contains temporal resolution along with the detected periodicities. We performed this analysis on each pixel of the two DS time series, calculating the time-averaged wavelet power contained in three period bands, centered at 3, 5, and 7\,min. By averaging the power over all periods in the corresponding band, we created two-dimensional power maps, which were discussed extensively in Papers I and II.

From the photospheric magnetic field values, we calculated the vector of the magnetic field up to 2000\,km with a potential field extrapolation (see Paper II). We calculated the inclination angle of the magnetic field vector with respect to the local vertical and the plasma-$\beta$ parameter, using the gas pressure values of the VAL C model (Vernazza \etal \cite{vernazza}). At each pixel of the FOV, the location of the magnetic canopy was determined as the height where $\beta$ is of order unity. More details on the extrapolation process and the three-dimensional configuration of the magnetic field can be found in Paper II. 

At the magnetic canopy, the Alfv\'{e}n and sound speeds are almost equal. When propagating through this layer, also termed ``equipartition layer'', magneto-acoustic waves undergo mode conversion/transmission. In the two-dimensional case, an acoustic wave that meets the equipartition layer will transfer its energy partly to a slow magnetoacoustic wave (a process termed transmission) and partly to a fast magnetoacoustic wave (termed conversion). The former is an acoustic wave that propagates along the magnetic field lines, and the latter a compressible magnetic wave that propagates across the magnetic field lines. Therefore, transmission preserves the nature of the wave, i.e. an acoustic wave in a high-$\beta$ environment (a fast acoustic wave), remains essentially acoustic in a low-$\beta$ plasma, which follows the direction of the magnetic field (slow magnetoacoustic). On the other hand, conversion preserves the identifier of the wave; that is, a fast high-$\beta$ acoustic wave remains fast, but changes nature, i.e. from acoustic to magnetic, in the low-$\beta$ plasma. Using generalized ray theory, Cally (2006) shows that the amount of energy transferred from the fast acoustic to the slow magnetoacoustic mode is defined by the transmission coefficient $T$ as

\begin{equation}
T=exp(-\pi k h_{s} sin^{2} \alpha)
\label{eq:t_coeff}
\end{equation}

\noindent where $k$ is the wavevector, $h_{s}$  the width of the equipartition layer (as seen by the wave), and $\alpha$  the angle between the wavevector and the magnetic field, which is the attack angle. Consequently, the conversion coefficient is given by 

\begin{equation}
C=1-T
\label{eq:c_coeff}
\end{equation}

\noindent as required by wave energy conservation. We should note that this formulation does not consider the existence of an acoustic cut-off. However, it has been shown by Schunker \& Cally (\cite{schunker06}) that the general conclusions drawn from this formulation are not affected. Furthermore, Cally (\cite{cally07}) shows that the position of maximum transmission and the shape of the $T$ curve follow those given by a wave mechanical calculation very closely. That being said, we proceed to the calculation of $T$ assuming that the wavenumber is given by $k =2\pi v / c_{s}$, where $v$ is the wave frequency, $c_{s} = \sqrt{\gamma P/ \rho}$ is the acoustic speed, $\gamma$=5/3 the adiabatic constant, $P$ the gas pressure, and $\rho$ the density (with the last two taken from the VAL C model). For simplicity, we assume that, below the canopy, waves propagate vertically upward, and therefore, the attack angle coincides with the inclination angle, $\theta$, of the magnetic field at the canopy height, estimated by the magnetic field extrapolation. Furthermore, we assume that $h_{s}$ is equal to the height resolution of the magnetic field extrapolation (235\,km).   

Slow waves propagate along the slanted magnetic field lines and escape the chromosphere as long as their frequency is higher than the modified acoustic cut-off, i.e., the cut-off reduced by the factor cos$\theta$  (ramp effect), $\theta$ being the magnetic field inclination to the vertical. Projection effects and the ramp effect should be taken into account in order to compare the transmission coefficient with the observations. To this end, we follow the approach proposed by Stangalini \etal (\cite{stang11}) and calculate the modified transmission coefficient:

\begin{equation}
T'=T(1-exp(-\frac{\theta}{\theta_{c}})) cos\theta 
\label{eq:t_stang}
\end{equation}

\noindent where $cos\theta$ compensates for the projection effects for slanted propagation of slow waves along the magnetic field lines, and $\theta_{c}$ is the critical inclination angle of the magnetic field vector over which propagation is allowed for waves with frequency lower than the acoustic cut-off. The second factor was selected by Stangalini (personal communication) to simulate p-mode leakage and resembles a step function, but with a smooth transition from non-transmission to transmission. In the presence of inclined magnetic field, the acoustic cut-off frequency is $v=v_{0}cos\theta$ (where $v_{0}$=5.2\,mHz), and the inversion of this formula gives $\theta_{c}$. We should note here that in an active region, where the equipartition layer is situated below the chromosphere, the ramp effect ensures that the transmitted slow waves will continue to propagate higher and, therefore, be observed. In the quiet Sun, on the other hand, the equipartition layer is situated well above the photosphere, and thus the ramp effect allows long-period waves to escape the photosphere and reach the magnetic canopy. Above the magnetic canopy, the acoustic cut-off is lowered because of both the inclined magnetic field and the lower sound speed. Similarly, for conversion, the ramp effect and projection effects may be accounted for by using the modified conversion coefficient

\begin{equation}
C'=C(1-exp(-\frac{\theta}{\theta_{c}})) cos(90\degr-\theta) 
\label{eq:c_stang}
\end{equation}

\noindent where the cos(90\degr-$\theta$) factor is the correction for projection effects, for propagation vertically to the magnetic field. 

Unlike slow waves, fast waves refract and propagate downward after meeting layers with a large fast-speed gradient. Although fast waves should follow a direction perpendicular to the magnetic field, simulations indicate that the magnetic field direction does not restrict the propagation path (Nutto \etal \cite{nutto12a}). Instead, it is the sharp gradient of the phase speed that leads to the refraction of the fast waves, which are then directed downward. A simple approach to estimating the turning height, i.e. the height at the chromosphere where fast waves undergo total internal reflection, is to consider the refraction of fast waves on an interface between two media where the phase speeds are different. According to Snell's law for refraction,

\begin{equation}
\frac{sin\theta_{i}}{sin\theta_{r}}=\frac{\upsilon_{i}}{\upsilon_{r}}
\label{eq:snell2}
\end{equation}

\noindent where  $\theta_{i}$ and $\theta_{r}$ are the incidence and refraction angles, and $\upsilon_{i}$ and $\upsilon_{r}$ are the phase speeds of the incident and the refracted waves, respectively. We assume that the incident wave at the equipartition layer is an acoustic wave with phase speed equal to the local sound speed, while the refracted wave is the fast magnetoacoustic wave with a phase speed equal to the speed of the fast wave. Taking this into account, the previous relation becomes

\begin{equation}
sin\theta_{r}=\frac{\upsilon_{f}}{c_{s}}sin\theta_{i}
\label{eq:refr1}
\end{equation}

\noindent where

\begin{equation}
\upsilon_{f}=\sqrt{\frac{1}{2}(\upsilon_{A}^{2}+c_{s}^{2})
+\frac{1}{2}\sqrt{(\upsilon_{A}^{2}+c_{s}^{2})^{2}-4\upsilon_{A}^{2}c_{s}^{2}cos^{2}\theta}}
\label{eq:u_fast}
,\end{equation}

\noindent and

\begin{equation}
\upsilon_{A}=\frac{B}{\sqrt{\mu \rho}}
\label{eq:alfven}
\end{equation}

\noindent is the Alfv\'{e}n speed, with $\mu$  the magnetic permeability of vacuum (4$\pi$ 10$^{-7}$ N/A$^{2}$).

Total internal reflection occurs when $sin\theta_{r}$ becomes greater than unity, which is equivalent to $\theta_{r}$ assuming imaginary values. In vertical propagation, the incidence angle is defined by the inclination of the magnetic field at the height of refraction.  Therefore, $\theta_{i}$=90\textdegree - $\theta$, where $\theta$ is the local inclination of the magnetic field, and the condition for total internal reflection becomes

\begin{equation}
\frac{\upsilon_{f}}{c_{s}}cos\theta > 1
\label{eq:refr_final}
.\end{equation}

\noindent Using this relation we can estimate the turning height as the minimum height where the above condition is satisfied.

\section{Results}

\subsection{Magnetic field configuration and turning height}

As seen in the lefthand panel of Fig.~\ref{fig:data}, the FOV of our observations contains a well formed rosette of mottles (lower right), as well as a smaller one (upper right). The protruding dark mottles surround the positive polarity magnetic concentrations (Fig.~\ref{fig:data}, right panel), which partly outline the network. In Paper II we showed that in this typical quiet solar area, mottles follow the chromospheric magnetic field and outline the magnetic flux tubes, some of which appear to connect the network with negative polarity magnetic fields at the internetwork and other network locations. They also form the magnetic canopy, which modifies the observed acoustic power. From Fig.\,5 of Paper I and Fig.\,4 of Paper II, it is evident that the 3 and 5\,min power is suppressed over the rosette at the chromosphere (magnetic shadow) and enhanced at the photosphere (power halo). We also showed in these papers that the height of formation of the magnetic canopy, relative to the height of formation (HOF) of the corresponding \ha bandpass, is responsible for the formation of both the magnetic shadow and the power halo: at the chromosphere above the magnetic canopy, we observe suppression of power, while at the photosphere, around the network and below the canopy, we find increased power. Since the magnetic canopy divides low-$\beta$ from high-$\beta$ regions, it has been made clear that it is imperative to know whether one observes above or below the canopy (Rosenthal \etal \cite{ros02}, Bogdan \etal \cite{bog03}, Finsterle \etal \cite{finsterle}). These and similar findings in other studies (see Sect.\,1) have been interpreted on grounds of the transmission/conversion of waves, which take place on the magnetic canopy. In the following, we investigate these processes further in comparison with our observations.

In the middle panel of Fig.~\ref{fig:3dplot} we have plotted the contours of the turning height, over the rosettes of Fig.~\ref{fig:data}. For comparison, we have also plotted the contours of the magnetic canopy height (Fig.~\ref{fig:3dplot}, left panel), calculated as in Paper II. The turning height is equal to the canopy height above the network bright points (magnetic conentrations), within the spatial resolution element, but it becomes progressively higher as the distance from the network increases. This is anticipated since, for small inclination angles (i.e., over the network), it can be shown after some algebra that Eq.~\ref{eq:refr_final} is equivalent to the definition of the magnetic canopy. Indeed, for small $\theta$, cos$\theta$ tends to unity and Eq.~\ref{eq:u_fast} becomes

\begin{equation*}
\upsilon_{f}=\sqrt{\frac{1}{2}(\upsilon_{A}^{2}+c_{s}^{2})
+\frac{1}{2}\sqrt{(\upsilon_{A}^{2}+c_{s}^{2})^{2}-4\upsilon_{A}^{2}c_{s}^{2}}} 
\end{equation*}
\begin{equation*}
=\sqrt{\frac{1}{2}(\upsilon_{A}^{2}+c_{s}^{2})
+\frac{1}{2}\sqrt{(\upsilon_{A}^{2}-c_{s}^{2})^{2}}}\\
\end{equation*}
\begin{equation*}
=\sqrt{\frac{1}{2}(\upsilon_{A}^{2}+c_{s}^{2})
+\frac{1}{2}(\upsilon_{A}^{2}-c_{s}^{2})}\\
\end{equation*}
\begin{equation*}
=\upsilon_{A}
\end{equation*}

\noindent which means that, for low inclination angles, Eq.~\ref{eq:refr_final} is equivalent to the condition for the transition from high to low-$\beta$ plasma and, therefore, gives the height of the magnetic canopy. The crucial finding here is that, for a wide range of inclination angles, these two layers do not coincide, and as a consequence, DSs from areas above the magnetic canopy (and below the turning height) may be due to fast waves as well. This finding is yet another reminder of how difficult the interpretation of velocity signals may be in terms of different magnetoacoustic modes, at the highly inhomogeneous chromospheric plasma.

This remark is illustrated vividly in the righthand panel of Fig.~\ref{fig:3dplot}. In this panel, a vertical slice of the fast speed along the horizontal dashed line in Fig.~\ref{fig:data} ($Y$ = 13$\arcsec$) is plotted, which crosses the network magnetic concentrations around $X = 27\arcsec$. As expected, the fast speed increases dramatically with height above the magnetic canopy and around the network, as the velocity contours in the righthand panel of Fig.~\ref{fig:3dplot} clearly indicate. It is these high fast-speed gradients that cause the reflection of fast waves (see, e.g., Nutto \etal \cite{nutto12a}), with the turning height situated above the magnetic canopy.

\begin{figure*}[htp]
\centering
  \includegraphics[width=20cm]{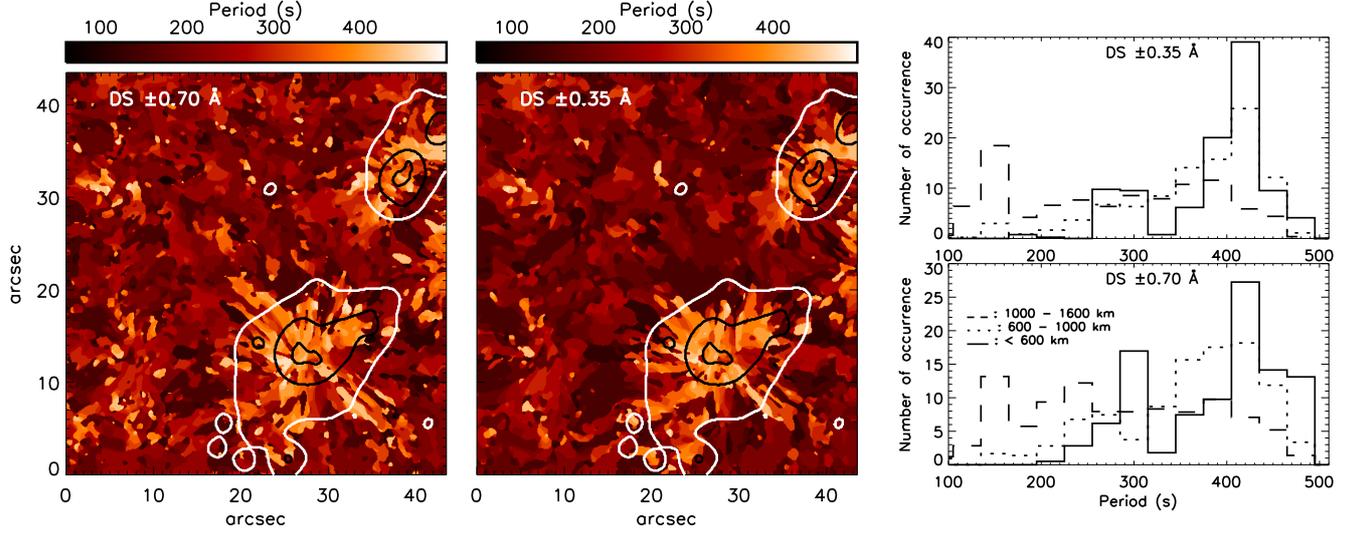}
  \caption[]
    {Dominant period maps of the photospheric (left) and chromospheric (middle) DS oscillations and the respective histograms (right). The contours on the maps represent turning heights 600, 1000 (black curves), and 1600\,km (white curves) from $\tau_{5000}=1$. The histograms are calculated for both maps, using 30\,s bins for the period, for the pixels of the large rosette where the turning height ranges from 0 -- 600\,km (solid line), 600 -- 1000\,km (dotted line), and 1000 -- 1600\,km (dashed line).}
    {\label{fig:domper}}
\end{figure*}

In the same panel we have also included information concerning the estimated range of the HOF of the \ha at $\pm$0.70\,\AA\ and $\pm$0.35\,\AA, based on Leenaarts \etal (\cite{leenaarts}). It must be noted that it is not easy to attribute specific geometrical heights of formation to a given bandpass ignoring the dynamical character of the solar atmosphere. Thus, these estimates do not take the presence of the rosette and the dynamical elongated structures that form it into account. We hereafter refer to the two DS, estimated from the \ha wing intensities at $\pm$0.70\,\AA\ and $\pm$0.35\,\AA as photospheric and chromospheric velocities, respectively, keeping in mind, however, that in the \ha rosette region, these two DS are directly related to the dark structures forming it. It is obvious that, not only does each DS contain information from different dynamic regimes, but also different physical conditions are found within the same HOF of a bandpass. The photospheric velocities are mostly related to the acoustically dominated plasma below the canopy, except for a very narrow strip just above the network, where the canopy is below 600\,km and almost coincides with the turning height. Inside this area we do not expect to see fast waves, but their reflection will cause increased acoustic power around the network. Indeed, in Paper II we found that the acoustic power in this bandpass is affected by wave reflection, as long as the magnetic canopy forms below 1200\,km. 

Although the magnetic canopy is more extended horizontally, at heights greater than 1000\,km, chromospheric velocities contain mixed signal, as well. Over a broad area around the network, regions both above and around the magnetic canopy are sampled. Given the location of the turning height there, we expect to see slow waves only between the parts of the blue line intersected by the HOF of the chromospheric velocities. Outside this area, the velocity purturbations might be due to either type of wave, but will predominantly be due to fast waves, since the slanted propagation of the slow waves will result in reduced power detected along the LOS.

The role of the turning height is evident in Fig.~\ref{fig:domper}, where its contours are overplotted on the maps of photospheric and chromospheric dominant periods, corresponding to the period where the power is maximum in each pixel. In general, period variations longer than 5\,min dominate in the elongated structures around the magnetic network, and it is clear that the 1600\,km contour largely encloses these areas. If these variations represent acoustic oscillations, they should be the signature of slow magnetoacoustic waves, while outside this contour, the dominant low periods (high frequencies) may be attributed to fast waves.

As seen in the histograms of Fig.~\ref{fig:domper}, there is a lack of short periods (up to 250\,s) and the clear existence of 300\,s and longer periods, evident in both DS, at those pixels of the rosette where the turning height is formed lower than 600\,km, which largely correponds to the network. For turning heights between 600\,-\,1000\,km,  the number of pixels dominated by short periods is increased, but these are still outnumbered. This effect is reversed when the turning height is even higher, between 1000\,-\,1600\,km. From these histograms, we infer a gradual decrease in the occurrence of dominant low period oscillations since the turning height is situated lower. This, of course, does not mean that there is complete absence of short period power. As shown in Papers I and II, there is non-negligible power at the 3\,min range at the network, but it seems that there is more power at longer period waves, indicating that a mechanism favors the propagation of these waves at the rosette, and is directly related to the elongated structures forming it, where the magnetic field is highly inclined.

An important remark should be made here concerning long-period oscillations: assuming that the DS at $\pm$0.70\,\AA\ has a photospheric origin, one would expect that the corresponding dominant period map should exhibit long periods only inside the 1000\,km or 600\,km contours, differing markedly from the chromospheric map. However, this is not the case, since we find elongated structures showing long periods, even more extended in the photosphere than in the chromosphere. These long periods could be attributed to the appearance and disappearance of Doppler-shifted features on the wavelet spectrum. In fact, these variations would affect the spectrum of both DS similarly, with a tendency to produce larger features on the $\pm$0.70\,\AA\ DS, where these structures are visible in higher contrast. A similar effect may be produced by long-period slow magnetoacoustic waves running along the low-lying magnetic flux tubes that connect the network with unresolved internetwork magnetic patches, smeared out by the resolution of our magnetogram.

\subsection{Transmission and conversion of waves on the magnetic canopy}

\begin{figure*}[htp]
\centering
  \includegraphics[width=12cm]{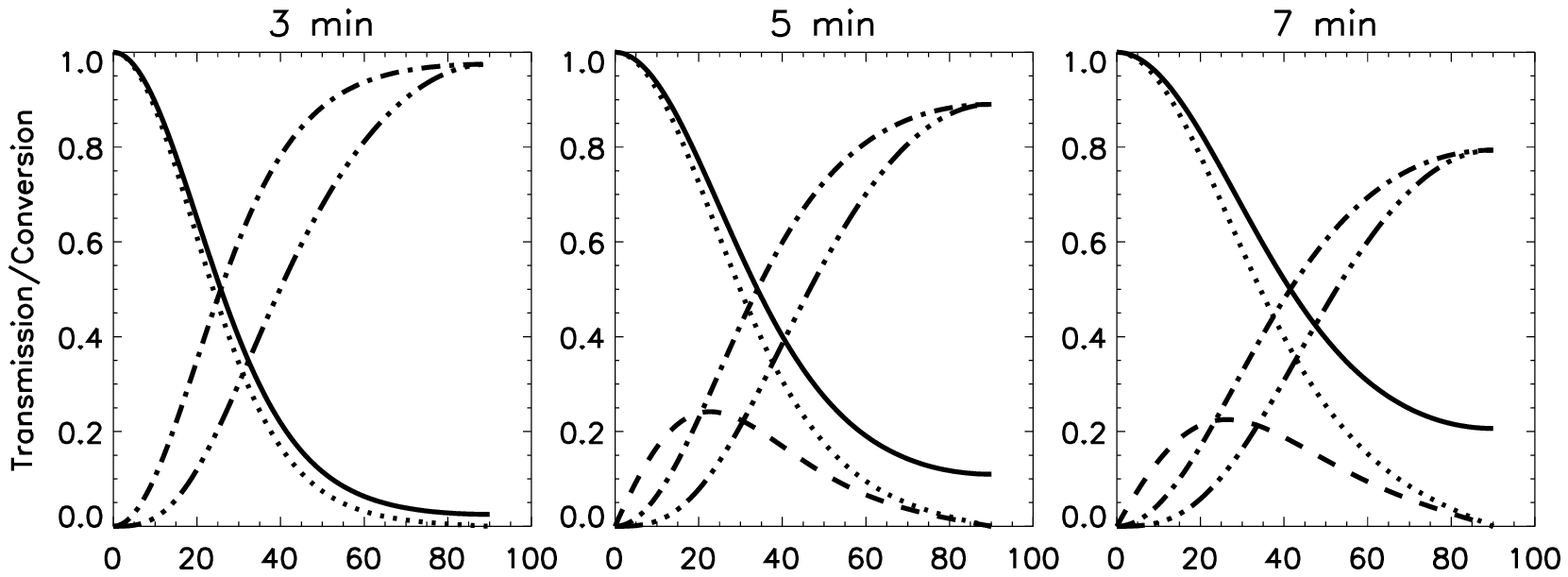}
  \includegraphics[width=12cm]{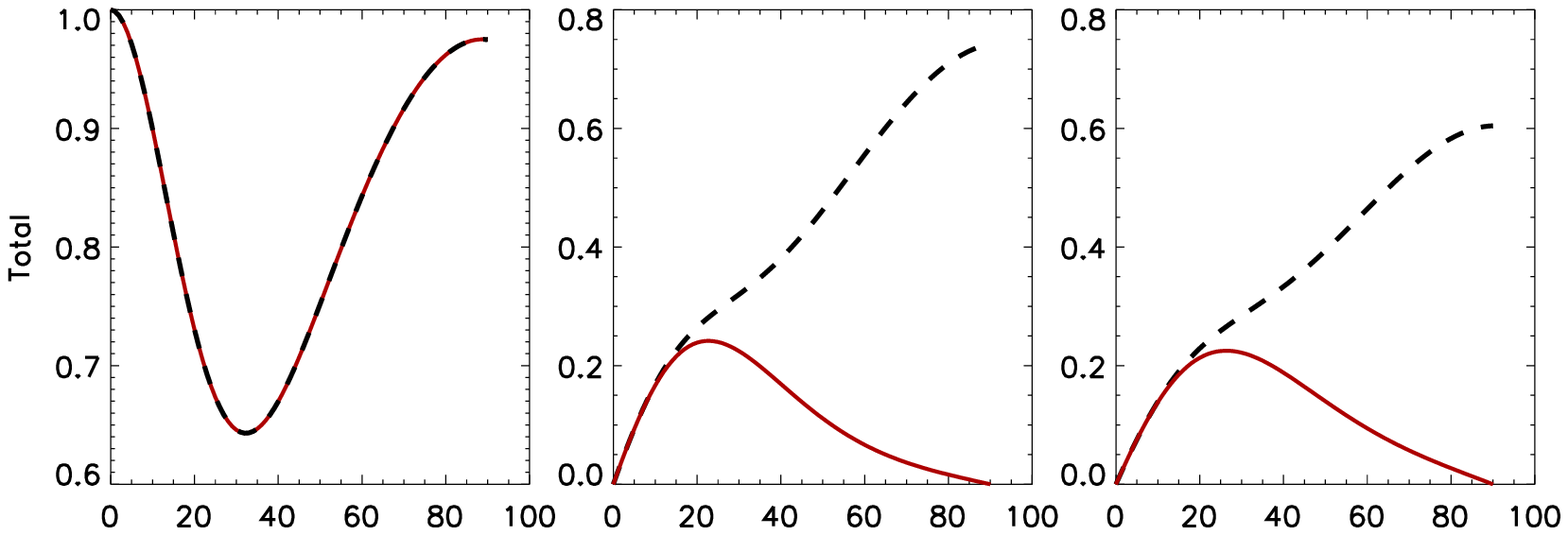}
  \includegraphics[width=12cm]{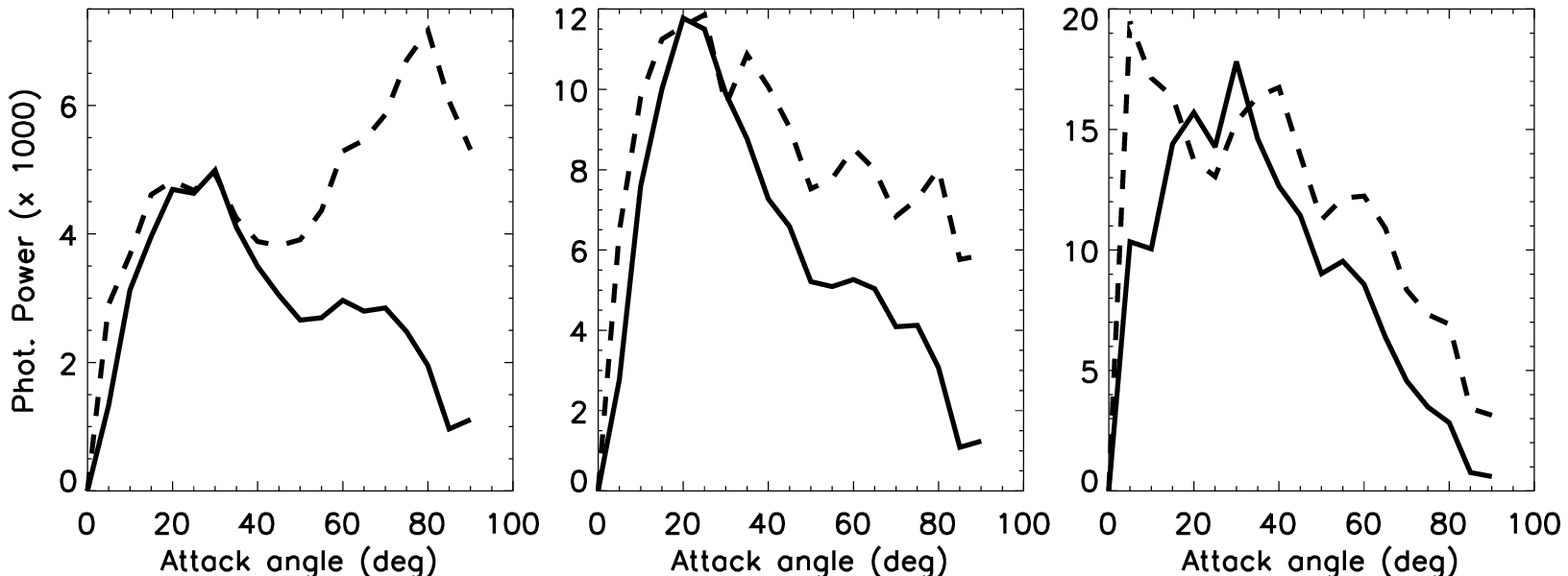}
\caption[]
    {First row: Transmission and conversion coefficients for 3, 5 and 7\,min waves (left, middle and right panels, respectively). The solid line represents $T$ (Eq.~\ref{eq:t_coeff}), while the dotted one $T$ corrected for projection effects ($T$ multiplied by $cos\theta$). The dashed-dotted line represents the conversion coefficient, $C=1-T$ and the dashed-dotted-dotted the one corrected for projection effects. In the middle and right panels, the dashed lines represent $T'$, which, in addition, takes into account the ramp effect (Eq.~\ref{eq:t_stang}). Second row: The dashed lines represent the sum of transmission and conversion, both corrected for projection and ramp effects. Red lines mark the acceptable curve for each period in quiet Sun (see text). Third row: Photospheric (solid line) and chromospheric (dashed line) power versus the inclination at the canopy (attack angle) for 3, 5 and 7\,min oscillations, calculated from observations.}
    {\label{fig:transm}}
\end{figure*}

In this section we try to qualitatively explain the transmission and conversion of magnetoacoustic waves on the magnetic canopy in relation to our observations. We use the two--dimensional model proposed by Schunker \& Cally (\cite{schunker06}) and Cally (\cite{cally07}) and the modifications proposed by Stangalini \etal (\cite{stang11}) (see Sect.\,3). In the first row of Fig.~\ref{fig:transm} we plot the coefficients $T$, $T'$, $C,$ and $C'$, as functions of the attack angle (magnetic field inclination at the $\beta=1$ layer), for the 3, 5, and 7\,min period bands. Also included are the coefficients $T$ and $C$, corrected only for projection effects. Especially, for the 3\,min waves, only the latter correction is necessary, since we consider that these waves may propagate upward regardless of the inclination of the magnetic field, and the ramp effect does not apply. We remind the reader that $T$ and $C$ express the percentage of acoustic power transferred to the slow and fast magnetoacoustic waves (respectively), if we assume that a propagating acoustic wave reaches the magnetic canopy from below. Similarly, $T'$ and $C'$ are modified versions of $T$ and $C$ (according to Sect.\,3) and express the percentage of the acoustic power one would observe along the canopy.

As clearly demonstrated in Fig.~\ref{fig:transm} (first row) and noted before, transmission is favored by small attack angles and long periods, showing a steeper decrease with attack angle for shorter periods and considerable, non-zero values for longer periods at high attack angles. The latter effect completely disappears when these curves are corrected for projection effects, as the transmitted (slow) waves follow a slanted propagation path. The ramp effect introduces further corrections to the 5 and 7\,min transmission curves (middle and right panels of Fig.~\ref{fig:transm}), with transmission finally becoming zero at  0$\degr$ and maximum around 20$\degr$--25$\degr$. This maximum is the combined result of both transmission decrease with attack angle and its increase as the attack angle approaches the critical angle (ramp effect).

On the other hand, the conversion coefficient (Fig.~\ref{fig:transm}, first row) increases with increasing attack angle. At 3\,min, for attack angles around $40\degr$, more than 80\% of the power is already transferred to the fast magnetoacoustic mode. For 5 and 7\,min waves, conversion is less efficient. Since fast waves should propagate vertically to the magnetic field, projection effects mostly affect (decrease) the values of the conversion coefficient at small attack angles (Fig.~\ref{fig:transm}, first row).

In practice, one cannot distinguish between transmission and conversion and 
as a result observed velocity signals cannot be interpreted on the grounds of one type of waves. To further complicate the picture, we should mention that there are numerous other types of waves supported by magnetic flux tubes embedded in a magnetohydrodynamic environment such as torsional waves, kink waves, etc. (see, e.g., Mathioudakis \etal \cite{math13}), which are, however, beyond the scope of the present study. To investigate the total anticipated acoustic power on the magnetic canopy, we synthesise transmission and conversion in a single curve for each period (second row of Fig.~\ref{fig:transm}), by adding the coefficients $T'$ and $C'$ for each period. Although, according to Eq.~\ref{eq:c_coeff} adding $T$ and $C$ should result to unity and, therefore, to constant total observed acoustic power with attack angle, this is not the case for the sum of $T'$ and $C'$. The combination of the different propagation directions of the slow and fast magnetoacoustic modes, parallel and vertical to the magnetic field, respectively, and of the ramp effect (for 5 and 7\,min waves only) finally results in the curves shown in the second row panels of Fig.~\ref{fig:transm}. The effect of the location of the HOF with respect to the canopy and turning heights on these curves has not been taken into account and  is discussed in detail later in this section, when we compare them to the power curves derived from observations. 

Waves with periods shorter than or close to 3\,min may freely propagate upward regardless of the attack angle. Therefore, increased observed power is expected at both low and high attack angles (Fig.~\ref{fig:transm}, second row, left panel), the former due to transmission (slow waves), while the latter due to conversion (fast waves). At intermediate attack angles, power is decreased as a result of the projection effects. For the 5 and 7\,min waves, the superposition of $T'$ and $C'$ gives the corresponding curves in Fig.~\ref{fig:transm} (second row, middle and right panels). Their shape differs remarkably from the one for the 3\,min waves, as they exhibit a continuous increase. As the attack angle increases, power increases due to the modified transmission, but also due to increasing conversion. 

\begin{figure}[h]
\centering
  \includegraphics[width=9cm]{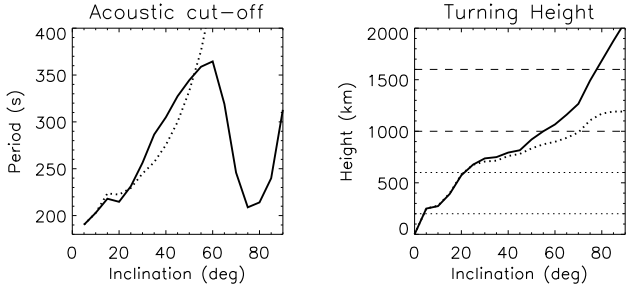}
  \caption[]
    {Left: acoustic cut-off period as function of the inclination of the magnetic field at the photosphere (solid line) and at the canopy level (dotted line). Right: turning height (solid line) and canopy height (dashed line) versus the atack angle (inclination at the canopy height). For reference, the estimated HOF (Leenaarts \etal \cite{leenaarts}) of photospheric ($\pm0.70\,\AA$ from line center) and chromospheric ($\pm0.35\,\AA$ from line center) \ha line wings are included (horizontal dotted and dashed lines, respectively).}
    {\label{fig:height_vs_incl}}
\end{figure}

For the construction of these two curves, it was assumed that 5 and 7\,min waves may also reach the magnetic canopy regardless of the magnetic field inclination. However, this is not the case in reality. From the inversion of $v=v_{0}cos\theta$, we find that 5 and 7\,min waves may propagate upward only for inclination angles larger than 50.6$\degr$ and 62.5$\degr$, respectively. In the lefthand panel of Fig.~\ref{fig:height_vs_incl}, we have measured the average cut-off period at the photosphere and at the canopy level contained in 5$\degr$ wide bins of attack angle. Above the magnetic canopy, the ramp effect ensures that, upon reaching the critical angle, the acoustic cut-off period will continue to increase, allowing the propagation of longer period waves. However, at the photosphere, the ramp effect has an important implication. As we move away from the network, although the inclination increases, the cut-off period decreases again because the magnetic field decreases. 

Long-period waves may reach the magnetic canopy only near and around the magnetic network by following the direction of the inclined magnetic field. Thus, at the equipartition layer the attack angle of these waves is zero, instead of being equal to the local inclination of the magnetic field. No conversion to fast magnetoacoustic waves is therefore expected for the long-period waves, and one expects to observe, instead, a power distribution similar to the dashed curves in the first row panels of Fig.~\ref{fig:transm}. It is expected that the ``escaping'' long-period waves will increase the power at the outer parts of the equipartition layer. As a consequence, in a typical quiet Sun configuration, as the one shown in Fig.~\ref{fig:3dplot}, the pixels that correspond to higher attack angles (higher inlination angles of the magnetic field) will contain more power than is predicted by $T'$. 

In the second row of Fig.~\ref{fig:transm} we have plotted the curves that describe the total (transmitted plus converted) power coefficient, given by the sum of the two corresponding coefficients for each period band, following the preceding analysis and discussion. With our data we have the opportunity to study the power of acoustic oscillations as a function of the attack angle, assumed to be equal to the inclination of the magnetic field at the height of the magnetic canopy. 

The magnetic canopy itself is a curved layer that intersects the HOF of the wings of \ha at different locations on the FOV. This is evident in the righthand panel of Fig.~\ref{fig:3dplot}. The previous section discussed that both DS contain contributions from diverse dynamic regions of the solar plasma. This is also illustrated in the righthand panel of Fig.~\ref{fig:height_vs_incl}, where we have plotted the average canopy height and turning height contained in 5$\degr$ wide bins of attack angle. As the attack angle increases progressively, both the turning height and the magnetic canopy height are situated higher, intersecting the HOF of the two DS at different heights. In this panel, the HOF of the \ha $\pm$0.70\,\AA\ and $\pm$0.35\,\AA\ from line center, estimated by Leenaarts \etal (\cite{leenaarts}) are also plotted for comparison. We have already shown that mottles are canopy structures (Paper II) and that at the rosette the HOF of the two DS are probably curved layers. Furthermore, DS at $\pm$0.35\,\AA\ is located higher than the DS at $\pm$0.70\,\AA, with the latter containing more photospheric signal. It is evident that, unlike the red curves in Fig.~\ref{fig:transm} (second row), which are calculated along the equipartition layer, the corresponding curves based on real data will contain mixed power from below- and above-canopy areas.

Nonetheless, we attempted to compare theory and observations and, to this end, constructed the observed photospheric and chromospheric power curves in the third row of Fig.~\ref{fig:transm}. We measured the average acoustic power of 3, 5, and 7\,min at the photosphere and the chromosphere, contained in $5\degr$ wide bins of attack angle on the pixels of the rosette region and, more specifically, on the part where the magnetic canopy forms lower than 1200\,km. At this part of the FOV, both the magnetic shadow (at the chromosphere) and the power halo (at the photosphere) are observed (Paper II). In Fig.~\ref{fig:3dplot} (right panel), this area lies between $X = 18\arcsec$ and $X =41\arcsec$. We avoided including a larger area because the amount of signal originating in the non-rosette photosphere and chromosphere will be increased, aliasing the dependence of power on the attack angle. It is quite remarkable that, regardless of the simplified assumptions and the complicated geometry, the shape of the power curves exhibit qualitative similarities to the corresponding constructed red curves shown in the panels of the second row in Fig.~\ref{fig:transm}.

The most striking feature of the 3\,min power curves (Fig.~\ref{fig:transm}, third row, left panel) is their two peaks, one for small (around 20$\degr$) and one for large attack angles (around 60$\degr$ and 80$\degr$ at the photosphere and chromosphere, respectively). An interesting feature, also, is that the maximum power at the photosphere is found at low attack angles, while the opposite happens at the chromosphere. As already noted, the first peak of the curve marks transmission to the slow magnetoacoustic mode at the photosphere. As these waves travel along the magnetic field lines and since the inclination of the magnetic field is higher at the chromosphere (see Paper II), the corresponding peak at the chromosphere is somewhat lower due to projection effects. On the other hand, high inclination favors conversion into fast magnetoacoustic waves at the magnetic canopy. (We note that the conversion coefficient is $C=1-T$, therefore conversion increases as transmission decreases.) As seen in the righthand panel of Fig.~\ref{fig:height_vs_incl}, for attack angles larger than 60$\degr$, the turning height is formed above 1000\,km and thus situated high enough so as to not interfere with the HOF of the \ha photospheric velocity. At the chromosphere, the turning height is higher than the corresponding HOF for attack angles larger than $80\degr$, and therefore the peak is shifted toward larger attack angles. Also, for very low attack angles, the turning height is lower than the photospheric HOF, and the 3\,min acoustic power is very low, so we do not expect any fast mode contribution to the oscillatory power. 

Although it is anticipated that acoustic power should be maximum at low attack angles, the observed power curves show the opposite. It should be noted that very few pixels correspond to low attack angles and even over the maximum magnetic flux (found at the center of the network cluster), the magnetic field is not exactly vertical. Thus, the sampling of power for low attack angles is statistically insignificant. Furthermore, the transmitted waves will not propagate vertically, but instead will be directed towards the outer parts of the rosette, and consequently, this area will be sampled at pixels corresponding to larger attack angles. 

Clearly, interpretation of 3\,min power is a complicated task since a) 3\,min acoustic waves may reach the chromosphere regardless of the magnetic field, b) they are easily converted to fast magnetic waves and may appear around the canopy at any height, and c) both slow and fast magnetoacoustic modes are detected as velocity oscillations. To distinguish between them it is necessary to take the curvature of the magnetic canopy and the position of the turning height into account, with respect to the HOF of the two \ha DS and the inclination of the magnetic field. 

It is easier to interpret 5\,min power (Fig.~\ref{fig:transm}, third row, middle panel), since (a) and (b), mentioned above, do not hold for this case. Photospheric and chromospheric 5\,min power generally follows the shape of the corresponding curves in the second row of Fig.~\ref{fig:transm}. Indeed, the photospheric power curve peaks at 20$\degr$, again as a result of increased transmission, and then decreases rapidly. As seen in Fig.~\ref{fig:height_vs_incl} (right panel), the canopy is found within the HOF range of the photospheric DS only up to 20$\degr$. As the canopy is situated progressively higher, the photospheric power drops quickly for angles greater than 20$\degr$. On the other hand, the chromospheric power curve is broader. As already noted, the chromospheric signal comes mostly from regions well above the magnetic canopy. There, as the magnetic field tends to become uniform, the dependence of the power on the attack angle is washed out. It is also interesting to note that the observed power curves show slightly increased power for attack angles larger than $50\degr$. We suggest that this is due to the increased transmission expected at attack angles larger than the critical angle, as already mentioned.

Similarly, 7\,min power at the photosphere (Fig.~\ref{fig:transm}, third row, right panel) largely follows the shape of the corresponding curve (in the overlying panel). The distribution is broader than the 5\,min one and peaks at 30$\degr$. However, the chromospheric power curve deviates substantially from the 5\,min one. With these remarks in mind and given that 5 and 7\,min waves should not differ in principle, we suggest that the majority of 7\,min power at the chromosphere is probably not related to acoustic oscillations. As also stated in the previous section, the lifetime of mottles may produce power at the 7\,min range and so may other types of Alfv\'{e}nic disturbances (Mc Ateer \etal \cite{mcateer02}, \cite{mcateer03}). Had it been otherwise, these oscillations should show similar behavior to 5\,min.

\section{Discussion and conclusions}

It has been suggested that the interaction of the acoustic waves with the magnetic canopy produces the magnetic shadow (McIntosh \etal \cite{mcintosh03}, Muglach \etal \cite{mug05}, Moretti \etal \cite{moretti}, Papers I and II). Upon reaching this critical layer from below, acoustic waves transfer their energy to the slow and fast magneto-acoustic modes. The former may continue their propagation along the slanted magnetic field lines, while the latter reflect and may increase the acoustic power at lower heights, forming the power halo (Khomenko \& Collados \cite{khom06}, \cite{khom09}). 

Mode transmission and conversion are often invoked to explain the distribution of the acoustic power on extended FOV in both the quiet Sun and around active regions, where the power halo and magnetic shadow phenomena occur (see Komm \etal \cite{komm} and Khomenko \& Calvo Santamaria \cite{khom13}). Numerical simulations by Nutto \etal (\cite{nutto12b}) show that the magnetic shadows in small scale magnetic fields are a result of such processes, while it has been shown that the reflection of the fast waves in active regions may produce power halos (Khomenko \& Collados, \cite{khom09}). In this study, we interpret, for the first time in the quiet Sun, the power variation in the magnetic shadow and power halo, comparing it with a model that quantifies these processes and brings out the key parameters of the mechanism.

Based on simple physical assumptions, we calculated, in a rosette region, the turning height, i.e. the height where fast waves reflect. We found that it coincides with the magnetic canopy only in the proximity of the magnetic network and is situated increasingly higher than the magnetic canopy as the distance from the magnetic network increases. As a consequence, some of the acoustic oscillations detected above the canopy may also be due to fast waves. This proved to be the case for part of the 3 min power at the rosette. We also found that the occurrence of dominant periods is affected by the position of the turning height. Short-period oscillations are dominant only in places where the turning height is situated higher than 1600\,km, i.e. higher than the HOF of the two \ha DS. 

The amount of energy transferred to each of these two modes depends strongly on the attack angle, the angle between the wave vector and the magnetic field. We used the simple analytical expression presented and discussed by Cally (\cite{cally06}, \cite{cally07}) and Schunker \& Cally (\cite{schunker06}). In these studies, transmission and conversion are quantified, through coefficients, $T$ and $C$, that express the amount of energy transmitted to the slow and fast magneto-acoustic modes, respectively. We also adopted and extended the modifications made by Stangalini et al. (\cite{stang11}), that compensate for the projection and ramp effects. The latter modification reintroduces the effect of the cut-off frequency, which was not taken into account in the original analytical expressions. We were therefore able to interpret the photospheric and chromospheric power variation as a function of the attack angle and the formation of the power halo and the magnetic shadow in terms of mode transmission and conversion, taking the following into account: 
\begin{itemize}
\item both fast and slow magneto-acoustic waves may be detected in velocity signals, 
\item both are subject to projection effects, 
\item long period waves may propagate upward and meet the magnetic canopy only in a certain area around the magnetic network, where the acoustic cut-off is appropriately lowered by the inclined magnetic field, 
\item these waves reach the magnetic canopy under zero attack angle and undergo transmission,
\item the curvature of the canopy and the turning height/layer produce deviations from the homogeneous case, and
\item each \ha DS intersects the magnetic canopy and the turning height at different locations so each samples different and diverse dynamic regimes.
\end{itemize}
 
Following our analysis, we conclude that the 3\,min power at the rosette is due to slow waves near the network, where the inclination of the attack angle favors transmission. At the periphery of the rosette, where these waves penetrate the magnetic canopy, we see the signature of fast waves. This was also portrayed in the dominant short periods outside the 1600\,km contour of the turning height in Fig.~\ref{fig:domper}. The fast waves reflect at the turning height, which is situated progressively higher than the magnetic canopy as the distance from the network increases. The power in 5\,min, on the other hand, is solely due to slow waves that propagate along the magnetic field lines. Fast long-period (5 and 7\,min) waves cannot be produced above the magnetic canopy, since these waves reach the chromosphere by following the inclined magnetic field and, therefore, transmit all their power to slow waves. In this sense, the magnetic field of the network acts as a directional filter, eventually allowing only velocity perturbations aligned with the local magnetic field vector in
the upper chromosphere (Cally \cite{cally07}). The 7\,min power at the chromosphere is difficult to interpret only in terms of slow magnetoacoustic waves. It is proved that, indeed, the magnetic shadow and the power halo phenomena owe their existence to the conversion and transmission of waves on the magnetic canopy, as assumed in our previous work. 

Our conclusions on the interaction between the magnetic field and the acoustic waves bring out an analogy between the magnetic field of the network and active regions. For example, Stangalini \etal (\cite{stang11}) interpret observations of an active region, suggesting the same mechanism. In that case, the magnetic canopy was formed entirely below the HOF of the Ca\,II 8542\,\AA\ line. Therefore, fast waves were not expected to reach the observed atmospheric layer due to their reflection at the canopy, and only slow waves contributed to the observed acoustic power. As a result, the power variation of 3 min oscillations as a function of the magnetic field inclination agreed very well with the variation of $T'$. In general, in active regions, where strong magnetic field concentrations are extended, it is expected that the homogeneous case is an adequate approximation to a larger extent than for the quiet Sun, where conditions vary dramatically within the same FOV. As already mentioned, owing to the curvature of the magnetic canopy, our DS time series sample areas of both high- and low--$\beta$, depending on the position on the FOV. Even though this complicates the interpretation of the results, quiet Sun observations give us the opportunity to study both conversion and transmission of waves on the magnetic canopy on the same FOV. Of course, attempting to fit the observed power curves with analytical expressions, in the manner of Stangalini \etal (\cite{stang11}), would require better knowledge of several parameters, such as exact formation heights and adequate scanning of the \Ha profile for the deduction of actual velocities instead of Doppler signals. The latter would also give the opportunity to estimate energy fluxes, providing valuable insight into the energy transported by waves. 

It has been made clear that the many processes that take place on the magnetic canopy by the upwardly propagating waves, i.e. fast/slow, transmission/conversion, and their dependence on the acoustic cut-off and the inclination of the magnetic field complicate the problem of energy transport. Some remarks can be made, however, about the efficiency of the two modes. i.e. fast and slow, to carry energy in the overlying solar layers. Slow waves are guided along the magnetic field lines up to the chromosphere and can escape upward or be reflected downward or even steepen into shocks releasing energy (Jefferies \etal \cite{jefferies06}). Fast waves, on the other hand, are refracted or totally reflected above the magnetic canopy at the so-called turning height. This would mean that a large part of the energy is returned back in the photosphere, making the mechanism of energy transport by fast modes ineffective. It has been shown, however, that in the three-dimensional problem, the fast mode converts partly to an Alfv\'{e}n wave before totally reflecting (Cally \& Goossens \cite{cally08}, Cally \& Hansen \cite{cally11}, Khomenko \& Cally \cite{khom11}, \cite{khom12}, Hansen \& Cally \cite{hansen12}, Felipe \cite{felipe12}). The analogy is clear: at the magnetic canopy, the fast and slow magneto-acoustic modes are coupled, while at the turning height, the fast magnetoacoustic wave is coupled to the Alfv\'{e}n wave. This coupling is not present in the two-dimensional case, but is an essential process since Alfv\'{e}n waves carry energy that may dissipate in the corona. These studies find that in many cases their flux exceeds that of the acoustic waves making Alfv\'{e}n waves potent carriers of non-thermal energy in the upper solar atmosphere (see review by Mathioudakis \etal \cite{math13}). In fact, studies of the quiet Sun have revealed Alfv\'{e}n waves in mottles and linked them with oscillations at the photospheric level (Jess \etal \cite{jess12}, Kuridze \etal \cite{kuridze13}), showing in the most prominent way, the importance of studying the coupling between different layers of the solar atmosphere.

To this aim, we must note the value of high resolution (spatial, temporal, and spectral) \ha observations combined with photospheric magnetic field observations. Recent advanced numerical simulations by Leenaarts \etal (\cite{leenaarts12}) on the formation of \ha have shown the importance of this line in the observations of the chromospheric magnetic field and confirmed our previous conclusion (Paper II) that mottles are low-$\beta$ structures representing the slanted flux tubes that form the magnetic canopy. Onto this ruffled (in reality) surface, mode transmission/conversion occurs and causes the magnetic shadow and power halo phenomena. An important future step would be to further link the \ha chromosphere with the overlying transition region and corona utilizing the state--of--the--art data of IRIS and SDO.

\begin{acknowledgements}
The observations have been funded by the Optical Infrared Coordination network (OPTICON, http://www.ing.iac.es/opticon), a major international collaboration supported by the Research Infrastructures Program of the European Commission''s sixth Framework Program. The research was partly funded through the project ``SOLAR-4068',' which is implemented under the ``ARISTEIA II'' Action of the  operational program ``Education and Lifelong Learning'' and is cofunded by the European Social Fund (ESF) and Greek national funds. The DOT was operated at the Spanish Observatorio del Roque de los Muchachos of the Instituto de Astrof\'{i}sica de Canarias. The authors thank P. S\"{u}tterlin for the DOT observations and R. Rutten for the data reduction. Hinode is a Japanese mission
developed and launched by ISAS/JAXA, collaborating with NAOJ as a domestic
partner, and NASA and STFC (UK) as international partners. Scientific operation
of the Hinode mission is conducted by the Hinode science team organized at
ISAS/JAXA. This team mainly consists of scientists from institutes in the partner countries. Support for the post-launch operation is provided by JAXA and
NAOJ (Japan), STFC (U.K.), NASA, ESA, and NSC (Norway). Hinode SOT/SP
Inversions were conducted at NCAR under the framework of the Community
Spectro-polarimetric Analysis Center (CSAC; http://www.csac.hao.ucar.edu).

\end{acknowledgements}

\end{document}